\documentclass[aps,prb,twocolumn,groupedaddress,showpacs]{revtex4-1} 
\usepackage{graphics}
\usepackage{graphicx}
\usepackage{amsmath}
\usepackage{amssymb}
\usepackage{amsthm}
\usepackage{epsfig}
\usepackage{dsfont}
\usepackage{xcolor} 
\usepackage[T1]{fontenc}
\usepackage{lmodern}
\begin{document}
\title{Thermal Stability of Superconductors}
\author{Jacob Szeftel$^1$}
\email[corresponding author :\quad]{jszeftel@lpqm.ens-cachan.fr}
\author{Nicolas Sandeau$^2$}
\author{Michel Abou Ghantous$^3$}
\author{Muhammad El-Saba$^4$}
\affiliation{$^1$ENS Cachan, LPQM, 61 avenue du Pr\'esident Wilson, 94230 Cachan, France}
\affiliation{$^2$Aix Marseille Univ, CNRS, Centrale Marseille, Institut Fresnel, F-13013 Marseille, France}
\affiliation{$^3$American University of Technology, AUT Halat, Highway, Lebanon}
\affiliation{$^4$Ain-Shams University, Cairo, Egypt}
\begin{abstract}
A stability criterion is worked out for the superconducting phase. The validity of a prerequisite, established previously for persistent currents, is thereby confirmed. Temperature dependence is given for the specific heat and concentration of superconducting electrons in the vicinity of the critical temperature $T_c$. The isotope effect, mediated by electron-phonon interaction and hyperfine coupling, is analyzed. Several experiments, intended at validating this analysis, are presented, including one giving access to the specific heat of high-$T_c$ compounds.
\end{abstract}
\maketitle
	\section{Introduction}
	In the mainstream view\cite{par,sch,tin}, the thermal properties of superconductors are discussed within the framework of the phenomenological equation by Ginzburg and Landau\cite{gin} (GL) and the BCS theory\cite{bcs}. However, since this work is aimed at accounting for the stability of the superconducting state with respect to the normal one, we shall develope an alternative approach, based on thermodynamics\cite{lan}, the properties of the Fermi gas\cite{ash} and recent results\cite{sz5,sz4}, claimed to be valid for \textit{all} superconductors, including low and high $T_c$ materials.\par 
	The outline is as follows : the specific heat of the superconducting phase is calculated in section $2$, which enables us to assess its binding energy and thereby to confirm and refine a necessary condition, established previously for the existence of persistent currents\cite{sz4}; section $3$ is concerned with the inter-electron coupling, mediated by the electron-phonon and hyperfine interactions; new experiments, dedicated at validating this analysis, are discussed in section $4$ and the results are summarised in the conclusion.
	\section{Binding Energy}
\begin{figure}
\includegraphics*[height=7 cm,width=7 cm]{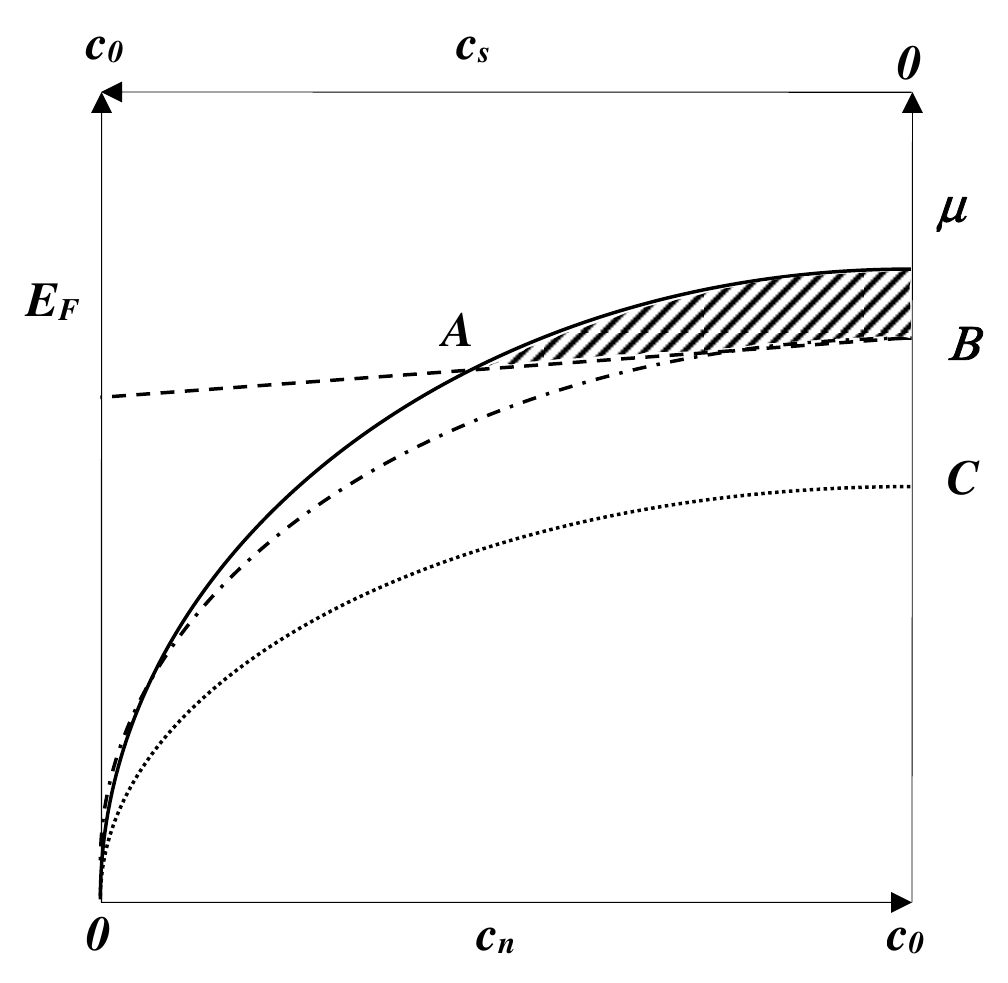}
\caption{Schematic plots of $E_F(T=0,c_n)$, $E_F(T_c,c_n)$, $E_F(T>T_c,c_n)$ and $\mu(c_s)$ as solid, dashed-dotted, dotted and dashed lines, respectively; $\frac{\partial \mu}{\partial c_s}$ has been taken to be constant for simplicity; the origin $E_F=\mu=0$ is set at the bottom of the conduction band; the crossing points $A,B$ of $E_F(0,c_n),E_F(T_c,c_n)$, respectively, with $\mu(c_s)$, exemplify stable solutions of Eq.(\ref{gidu}); the tiny differences $E_F(T,c_n)-\mu(c_0-c_n)$ have been hugely magnified for the reader's convenience.}
\label{sta3}
\end{figure}

	As in our previous work\cite{sz4,sz5,sz1,sz2,sz3,sz7}, the present analysis will proceed within the framework of the two-fluid model, for which the conduction electrons comprise bound and independent electrons, in respective temperature dependent concentration $c_s(T),c_n(T)$. They are organized, respectively, as a many bound electron\cite{sz5} (MBE) state, characterised by its chemical potential $\mu(c_s)$, and a Fermi gas\cite{ash} of Fermi energy $E_F(T,c_n)$. The Helmholz free energy of independent electrons per unit volume $F_n$ and $E_F$ on the one hand, and the eigenenergy per unit volume $\mathcal{E}_s(c_s)$ of bound electrons and $\mu$ on the other hand, are related\cite{ash,lan}, respectively, by $E_F=\frac{\partial F_n}{\partial c_n}$ and $\mu=\frac{\partial \mathcal{E}_s}{\partial c_s}$. At last, according to Gibbs and Duhem's law\cite{lan}, the two-fluid model fulfils\cite{sz4} at thermal equilibrium 
\begin{equation}
\label{gidu}
E_F(T,c_n(T))=\mu(c_s(T)),\quad c_0=c_s(T)+c_n(T),
\end{equation}
with $c_0$ being the total concentration of conduction electrons. Solutions of Eq.(\ref{gidu}) are given for $T=0,T_c$ in Fig.\ref{sta3}. Besides, Eq.(\ref{gidu}) has been shown\cite{sz5,sz4} to read for $T=T_c$ (see $B$ in Fig.\ref{sta3})
 \begin{equation}
\label{coo}
E_F(T_c,c_0)=\mu(c_s=0)=\varepsilon_c/2\quad , 
\end{equation} 
with $\varepsilon_c$ being the energy of a bound electron pair\cite{sz5}. Note that Eq.(\ref{coo}) is consistent with the superconducting transition, occuring at $T_c$, being of second order\cite{lan}, whereas it has been shown\cite{sz5} to be of first order at $T<T_c$, if the sample is flown through by a finite current. Then the binding energy of the superconducting state $E_b(T<T_c)$ has been worked\cite{sz5,gor} out as
\begin{equation}
\label{bin}
E_b(T)=\int_{T}^{T_c}\left(C_s(u)-C_n(u)\right)du\quad,  
\end{equation}
with $C_s(T),C_n(T)=\frac{\left(\pi k_B\right)^2}{3}\rho(E_F)T,$ being the electronic specific heat of a superconductor, flown through by a vanishing current\cite{sz5} and that of a degenerate Fermi gas\cite{ash} ($k_B,\rho(E_F)$ stand for Boltzmann's constant and the one-electron density of states at the Fermi energy). Due to Eq.(\ref{bin}), a stable superconducting phase $\Leftrightarrow E_b>0$ requires $C_s(T)>C_n(T)$, which is confirmed experimentally\cite{par,ash}, namely $C_s(T_c)\approx 3C_n(T_c)$.\par
	The bound and independent electrons contribute, respectively, 
	$$\begin{array}{l}
\mathcal{E}_s(c_s)=\int_{0}^{c_s}\mu(u)du\\
\mathcal{E}_n(T,c_n)=\int_{\epsilon_b}^{\epsilon_u}\epsilon\rho(\epsilon)f(\epsilon-E_F,T)d\epsilon
\end{array}\quad ,$$ 
to the total electronic energy $\mathcal{E}=\mathcal{E}_n+\mathcal{E}_s$. The symbols $\epsilon,f(\epsilon-E_F,T)$ refer to the one-electron energy and Fermi-Dirac distribution, while $\epsilon_b,\epsilon_u$ designate the lower and upper limits of the conduction band. Then, thanks to Eq.(\ref{gidu}) ($\Rightarrow dc_n+dc_s= dE_F-d\mu=0$), $C_s(T)=\frac{d\mathcal{E}}{dT}$ is inferred to read
\begin{equation}
\label{cs1}
C_s=\frac{\partial\mathcal{E}_n}{\partial T}-E_F\frac{\partial c_n}{\partial T}+\frac{dE_F}{dT}\left(\frac{\partial\mathcal{E}_n}{\partial E_F}-E_F\frac{\partial c_n}{\partial E_F}\right)
\quad,  
\end{equation}
with $c_n=c_n(T),c_s=c_s(T),E_F=E_F\left(T,c_n(T)\right)$. Because the independent electrons make up a degenerate Fermi gas ($\Rightarrow T<<T_F=E_F/k_B$), the following expressions can be obtained owing to the Sommerfeld expansion\cite{ash} up to $T^2$ 
\begin{equation}
\label{somm}
	\begin{array}{c} 
	\frac{\partial\mathcal{E}_n}{\partial E_F}=E_F\rho+\left(2\rho'+E_F\rho''\right)\frac{\left(\pi k_BT\right)^2}{6}\\ 
\frac{\partial\mathcal{E}_n}{\partial T}=	\left(\rho+E_F\rho'\right)\frac{\left(\pi k_B\right)^2}{3}T\\ 
\frac{\partial c_n}{\partial E_F}=\rho+\rho''\frac{\left(\pi k_BT\right)^2}{6}\quad ,\quad 
\frac{\partial c_n}{\partial T}=\rho'\frac{\left(\pi k_B\right)^2}{3}T
	\end{array}\quad ,
\end{equation}
with $\rho=\rho(E_F),\rho'=\frac{d\rho}{dE_F}(E_F),\rho''=\frac{d^2\rho}{dE_F^2}(E_F)$. Then Eq.\ref{cs1} is finally recast into
\begin{equation}
\label{cs2}
C_s(T)=\frac{\left(\pi k_B\right)^2}{3}\rho T\left(1+\frac{dE_F}{dT}\frac{\rho'}{\rho}T\right)
\quad.  
\end{equation}
Applying Eq.\ref{cs2} at $T=T_c$ yields
\begin{equation}
\label{ctc}
C_s(T_c)=C_n(T_c)\left(1+\frac{dE_F}{dT}(T_c^-)\frac{\rho'}{\rho}T_c\right)\quad .
\end{equation}
Hence it is in order to work out the expressions of $\frac{dE_F}{dT}(T>T_c)$ and $\frac{dE_F}{dT}(T\leq T_c)$.\par
	Due to $c_n(T>T_c)=c_0$, $\frac{dE_F}{dT}$ is deduced\cite{ash} to read 
\begin{equation}
\label{defp}
\frac{dE_F}{dT}(T>T_c)=-\frac{\frac{\partial c_n}{\partial T}}{\frac{\partial c_n}{\partial E_F}}=-\frac{\left(\pi k_B\right)^2}{3}\frac{\rho'}{\rho}T\quad ,
\end{equation}
which is integrated with respect to $T$ to yield 
\begin{equation}
\label{eft}
E_F(T=0,c_0)-E_F(T,c_0)=\frac{\left(\pi k_B\right)^2}{6}\frac{\rho'}{\rho}T^2\quad .
\end{equation}
Then consistency with Fig.\ref{sta3} requires $\rho'(E_F)>0$ so that $C$ goes toward $B$ for $T\searrow T_c$. Assuming $\rho(\epsilon)=\rho_f(\epsilon)\propto\sqrt{\epsilon}\Rightarrow\rho_f'(\epsilon)>0,\forall\epsilon$, with $\rho_f(\epsilon)$ being the density of states of three-dimensional free electrons, leads to 
	$$1-\frac{E_F(T>T_c)}{E_F(0,c_0)}=\frac{\pi^2}{12}\left(\frac{T}{T_F}\right)^2\quad .$$
A numerical application with a typical value $T_F=3\times 10^4K$ yields $T_F(300K)-T_F(0)\approx 3K<<T_F\Rightarrow\left|\frac{dT_F}{dT}\left(T>T_c\right)\right|<<1$.\par
	Taking advantage of Eq.\ref{gidu}, the expression of $\frac{dE_F}{dT}(T\leq T_c)$ is obtained to read
\begin{equation}
\label{fer}
\left.
\begin{array}{l}
dc_n=\frac{\partial c_n}{\partial E_F}dE_F+\frac{\partial c_n}{\partial T}dT\\
dc_s=\frac{\partial c_s}{\partial\mu}d\mu=-dc_n
\end{array}
\right\}\Rightarrow
 \frac{dE_F}{dT}=-\frac{\frac{\partial c_n}{\partial T}}{\beta(T)}\quad ,
\end{equation}
with $\beta(T)=\frac{\partial c_n}{\partial E_F}+\frac{\partial c_s}{\partial\mu}$. The Sommerfeld expansion (see Eq.(\ref{somm})) leads to 
\begin{equation}
\label{atc}
\alpha=\frac{dE_F}{dT}(T_c^-)\frac{\rho'}{\rho}T_c=-\frac{\left(\pi k_B\rho' T_c\right)^2}{3\rho\beta(T_c)} \quad .
\end{equation}
Thus, looking back at Eq.\ref{ctc}, it is realized that the observed\cite{par,ash} relation $C_s(T_c)\approx 3C_n(T_c)$ requires $\alpha>0\Rightarrow\beta(T_c)<0$, which had been already identified\cite{sz4} as a necessary condition for the superconducting state to be at thermal equilibrium. At last, $\alpha$ reads in case of $\rho=\rho_f$
	$$\alpha=\frac{\pi^2}{12}\left(\frac{T}{T_F}\right)^2\rho\frac{\partial E_F}{\partial c_n}\left(1+\frac{\partial E_F}{\partial c_n}\frac{\partial c_s}{\partial\mu}\right)^{-1}\quad .$$
Due to $\frac{T}{T_F}<<1$ and $\rho\frac{\partial E_F}{\partial c_n}\approx 1$, getting $\alpha\approx 2$ requires $\beta(T_c)\approx 0\Rightarrow\frac{\partial E_F}{\partial c_n}+\frac{\partial\mu}{\partial c_s}\approx 0$, so that the stability criterion of the superconducting state reads finally 
\begin{equation}
\label{cri}
\frac{\partial E_F}{\partial c_n}(T_c,c_0)=-\frac{\partial\mu}{\partial c_s}(0),\quad\rho'(E_F(T_c,c_0))>0\quad.
\end{equation}
Because of $\frac{\partial E_F}{\partial c_n}(T_c,c_0)\approx\frac{1}{\rho}>0$, Eq.(\ref{cri}) is seen to be consistent with $\frac{\partial\mu}{\partial c_s}(c_s)<0$, established previously as a prerequisite for persistent currents\cite{sz4} and the Josephson effect\cite{sz6}. At last, note that there is $\frac{dT_F}{dT}(T\leq T_c)>>1$ but inversely $0<-\frac{dT_F}{dT}(T>T_c)<<1$.\par
		In order to grasp the significance of the constraint expressed by Eq.(\ref{cri}), let us elaborate the case for which Eq.(\ref{cri}) is not fulfilled ($\Rightarrow C_s(T<T_c)<C_n(T)$). Accordingly the hatched area in Fig.\ref{sta3} is equal to the difference in free energy at $T=0$ between the superconducting phase and the normal one, and thence also equal to $E_b(0)>0$ because the entropy of the normal state vanishes\cite{lan} at $T=0$. However applying Eq.(\ref{bin}) with $C_s(T<T_c)<C_n(T)$ yields $E_b(0)<0$, which contradicts the above opposite conclusion $E_b(0)>0$, and thereby entails that the MBE state, associated with $A$ in Fig.\ref{sta3}, is \textit{not} observable at thermal equilibrium in case of \textit{unfulfilled} Eq.(\ref{cri}), even though it \textit{is} definitely a MBE eigenstate\cite{ja1,ja2,ja3} of the Hubbard Hamiltonian, accounting for the motion of correlated electrons, and its energy is indeed \textit{lower} than that of the Fermi gas $\mathcal{E}_n(T=0,c_0)$.\par
	Since energy and free energy are equal\cite{lan} at $T=0$, $E_b(0)$ reads
				$$E_b(0)=\int_{0}^{c_s(0)}\left(E_F(0,c_0-c_s)-\mu(c_s)\right)dc_s\quad .$$
In order to work out an upper bound for $E_b(0)$, $E_F(T,c_0-c_s)-\mu(c_s)$ will be approximated by its Taylor expansion at first order with respect to $c_s-c_s(T)$, which yields
\begin{equation}
\label{efmu}
	E_F(T,c_0-c_s)-\mu(c_s)\approx\frac{m}{e^2}\gamma\left(c_s(T)-c_s\right)\quad ,
\end{equation}				
with $\gamma=\frac{\partial E_F}{\partial c_n}(c_n(T))+\frac{\partial \mu}{\partial c_s}(c_s(T))$. Since it has been shown\cite{sz5} that $E_F(T,c_0-c_s)-\mu(c_s)<10^{-5}eV$, Eq.(\ref{efmu}) turns out to be very accurate. Likewise, due to $c_s(T)\geq c_s\geq 0$ (see Fig.\ref{sta3}) and $\gamma>0$ being a necessary condition\cite{sz4} for $A$ in Fig.\ref{sta3} to correspond to a stable equilibrium, Eq.(\ref{efmu}) entails
$$E_F(T,c_0-c_s)-\mu(c_s)\leq E_F(T,c_0)-\mu(0)\quad .$$
Then by taking advantage of Eqs.(\ref{coo},\ref{eft}), we get
$$E_F(T,c_0)-\mu(0)\leq E_F(0,c_0)-E_F(T_c,c_0)=\frac{\left(\pi k_BT_c\right)^2}{6}\frac{\rho'}{\rho},$$
with $\rho'>0$ as required by Eq.(\ref{cri}). At last, assuming $\rho(\epsilon)=\rho_f(\epsilon)$, the searched upper bound per electron is obtained to read
	$$\frac{E_b(0)}{c_0E_F(T_c,c_0)}\leq \frac{\pi^2}{12}\left(\frac{T_c}{T_F}\right)^2\quad .$$
Applying this formula to $Al$ ($T_c=1.2K,T_F\approx 3\times 10^4K$) gives $\frac{E_b(0)}{c_0E_F(T_c,c_0)}<10^{-8}$. Moreover, that latter result had enabled us to realize\cite{sz2} that the formula $E_b(0)=\mu_0H_c(0)^2/2$, albeit ubiquitous in textbooks\cite{par,sch,tin} ($H_c(T\leq T_c),\mu_0$ refer to the critical magnetic field and the magnetic permeability of vacuum, respectively), underestimates $E_b(0)$ by \textit{ten} orders of magnitude.\par
	Since fulfilling Eq.(\ref{cri}) is tantamount to $\beta(T_c)=0$, which entails  $\frac{dE_F}{dT}(T\rightarrow T_c^-)\rightarrow\infty$ and thence $C_s(T\rightarrow T_c^-)\rightarrow\infty$, it must be checked that $\mathcal{E}(T)=\int_0^T C_s(u)du$ remains still finite for $T\rightarrow T_c^-$. To that end, let us work out the Taylor expansion of $\mu(c_s),E_F(T,c_n)$ up to the second order around $T=0,c_s=0$
	$$\begin{array}{l}
\mu(c_s)=\mu(0)+\frac{\partial \mu}{\partial c_s}(0)c_s+\frac{\partial^2\mu}{\partial c_s^2}(0)	\frac{c_s^2}{2}\\
E_F(T,c_n)=E_F(T_c,c_0)-\frac{c_s}{\rho}-\frac{\rho'}{\rho^3}\frac{c_s^2}{2}\\
\quad\quad\quad\quad\quad +\frac{\left(\pi k_B\right)^2}{6}\frac{\rho'}{\rho}\left(T_c^2-T^2\right)\end{array}\quad ,$$
for which we have used $c_n=c_0-c_s,c_s=c_s(T),\frac{\partial E_F}{\partial c_n}=\frac{1}{\rho}\Rightarrow\frac{\partial^2 E_F}{\partial c_n^2}=-\frac{\rho'}{\rho^3}$. Then taking advantage of Eqs.(\ref{gidu},\ref{coo}) ($\Rightarrow E_F(T,c_n)=\mu(c_s),E_F(T_c,c_0)=\mu(0)$) and Eq.(\ref{cri}) ($\Rightarrow\beta(T_c)=\frac{\partial \mu}{\partial c_s}(0)+\frac{1}{\rho}=0$) results into
$$c_s(T\rightarrow T_c^-)=\pi k_B\sqrt{\frac{\rho'\left(T_c^2-T^2\right)}{3\left(\rho\frac{\partial^2\mu}{\partial c_s^2}(0)+\frac{\rho'}{\rho^2}\right)}}\propto\sqrt{T_c-T}
\quad .$$
It should be noticed that the GL equation predicts\cite{tin} rather $c_s(T\rightarrow T_c^-)\propto T_c-T$. \par
	Likewise, let us calculate similarly the Taylor expansion of $\beta(T)\propto\frac{\partial E_F}{\partial c_n}+\frac{\partial\mu}{\partial c_s}$ up to the first order around $T=T_c,c_s=0$  
	$$\begin{array}{c}
	\beta(T\rightarrow T_c^-)\propto\left(\frac{\partial^2\mu}{\partial c_s^2}(0)-\frac{\partial^2 E_F}{\partial c_n^2}(T_c,c_0)\right)c_s\Rightarrow\\
	\beta(T\rightarrow T_c^-)\propto\sqrt{T_c-T}\Rightarrow\mathcal{E}(T\rightarrow T_c^-)\propto\sqrt{T_c-T}
	\end{array}\quad ,$$
whence $\mathcal{E}(T\rightarrow T_c^-)$ is concluded to remain indeed finite.\par
	At last, we shall work out the expression of $j_M(T\rightarrow T_c^-)$, the maximum current density $j_s$, conveyed by bound electrons which was shown\cite{sz5} to read
$$j_M=ec_m(T)\sqrt{\frac{2}{m}\left(E_F\left(T,c_0-c_m(T)\right)-\mu(c_m(T))\right)}\quad ,$$
with $e,m$ standing for the charge and effective mass of the electron, while $c_m(T)=\frac{2}{3}c_s(T)$ designates the corresponding value of $c_s$, i.e. $j_s(c_m)=j_M$. Hence $j_M$ reads\cite{sz5} finally
$$\begin{array}{c}
j_M(T)=\frac{er}{\sqrt{m}}\left(\frac{2}{3}c_s(T)\right)^{1.5}\\
r=\sqrt{\frac{\partial E_F}{\partial c_n}(c_n(T))+\frac{\partial \mu}{\partial c_s}(c_s(T))}
\end{array}\quad .$$
It ensues from $\beta(T_c)=0$ that the leading term of the Taylor expansion of $r$ around $T=T_c,c_s=0$ reads
		$$\begin{array}{c}
		r\left(T\rightarrow T_c^-\right)\propto \sqrt{c_s(T)}\Rightarrow r\propto \left(T_c-T\right)^{\frac{1}{4}}\Rightarrow\\
		j_M\left(T\rightarrow T_c^-\right)\propto T_c-T
		\end{array}\quad ,$$
which is to be compared with the maximum persistent current density\cite{sz5} $j_c\left(T\rightarrow T_*^-\right)\propto\sqrt{T_*-T}$ with $T_*<T_c$.
				\section{Isotope Effect}
Substituting, in a superconducting material, an atomic species of mass $M$ by an isotope, is well-known\cite{par,sch,tin} to alter $T_c$. This isotope effect was ascribed to the electron-phonon coupling, on the basis of the observed relation $T_c\sqrt{M}=constant$. The ensueing theoretical treatment\cite{par,sch,tin} capitalised\cite{fro} on Froehlich's perturbation\cite{lan2} calculation of the self-energy of an independent electron induced by the electron-phonon coupling. However since the BCS picture\cite{bcs} has subsequently ascertained the paramount role of inter-electron coupling, we shall rather focus hereafter on the \textit{effective} phonon-mediated interaction between \textit{two} electrons. \par
	Thus let us consider independent electrons of spin $\sigma=\pm1/2$, moving in a three-dimensional crystal, containing $N$ sites. The  dispersion of the one-electron band reads $\epsilon(k)$ with $\epsilon(k),k$ being the electron, spin-independent ($\Rightarrow \epsilon(-k)=\epsilon(k)$) energy and a vector of the Brillouin zone, respectively. Their motion is governed, in momentum space, by the Hamiltonian $H_d$ 
	$$H_d=\sum_{k,\sigma}\epsilon(k)c^+_{k,\sigma}c_{k,\sigma}\quad ,$$ 
with the sum over $k$ to be carried out over the whole Brillouin zone. Then the $c^+_{k,\sigma},c_{k,\sigma}$'s are Fermi-like creation and annihilation operators\cite{sch} on the Bloch state $\left|k,\sigma\right\rangle$ 
	$$\left|k,\sigma\right\rangle=c^+_{k,\sigma}\left|0\right\rangle\quad ,\quad \left|0\right\rangle=c_{k,\sigma}\left|k,\sigma\right\rangle\quad ,$$
with $\left|0\right\rangle$ being the no electron state. Let us introduce now the electron-phonon\cite{par,sch,tin,fro} coupling $h_{e-\phi}$
	$$h_{e-\phi}=\frac{g_q}{\sqrt{N}}\sum_{k,k',\sigma}c^+_{k,\sigma}c_{k',\sigma}\left(a^+_q+a_{-q}\right)\quad ,$$ 
with $q=k'-k$ and $g_q\propto \left(\omega_qM\right)^{-1/2}$ being the coupling constant characterising the electron-phonon interaction. Likewise, $\omega_q$ is the phonon frequency, while the $a^+_q,a_{q}$'s are Bose-like creation and annihilation operators\cite{sch} on the $n_{q}\in\mathcal{N}$ phonon state $\left|n_{q}\right\rangle$
	$$a^+_q\left|n_q\right\rangle=\sqrt{n_q+1}\left|n_q+1\right\rangle \quad,\quad a_{q}\left|n_q\right\rangle=\sqrt{n_q}\left|n_q-1\right\rangle .$$\par
	Because of $\left\langle k\left|h_{e-\phi}\right|k'\right\rangle=0,\forall k,k'$ with $\left.|k\right\rangle=c^+_{k,+}c^+_{-k,-}\left.|0\right\rangle,\left.|k'\right\rangle=c^+_{k',+}c^+_{-k',-}\left.|0\right\rangle$, we shall deal with $h_{e-\phi}$ as a perturbation with respect to $H_d$, in order to reckon $\left\langle k\left|k'_2\right.\right\rangle$ with $\left|k'_2\right\rangle$ denoting $\left|k'\right\rangle$ perturbed at second order\cite{lan2}. Accordingly, we first introduce the unperturbed electron-phonon eigenstates
	$$\left|\widetilde{k}\right\rangle=\left|k\right\rangle\otimes\frac{\left|n_{q}\right\rangle+\left|n_{-q}\right\rangle}{\sqrt{2}},\quad 
\left|\widetilde{k'}\right\rangle=\left|k'\right\rangle\otimes\frac{\left|n_{q}\right\rangle+\left|n_{-q}\right\rangle}{\sqrt{2}}\quad ,$$
with $n_{q}=n_{-q}=n$. Their respective energies read $E(k)=2\epsilon(k)+n\hbar\omega_q$, $E(k')=2\epsilon(k')+n\hbar\omega_q$. Then we reckon $\left|\widetilde{k'_2}\right\rangle$ and further project it onto $\left|\widetilde{k}\right\rangle$, which yields
	$$\begin{array}{l}
	\left\langle\widetilde{k}\left|\widetilde{k'_2}\right.\right\rangle=\frac{g^2_q}{2N}\left(\left\langle\widetilde{k}
	\left|h_{e-\phi}\right|\varphi_+\right\rangle\left\langle\varphi_+\left|h_{e-\phi}\right|\widetilde{k'}\right\rangle\right.\\
	\quad\quad\quad\quad\quad\quad\left.+\left\langle\widetilde{k}
	\left|h_{e-\phi}\right|\varphi_-\right\rangle\left\langle\varphi_-\left|h_{e-\phi}\right|\widetilde{k'}\right\rangle\right)\\
	\varphi_+=c^+_{k,+}c^+_{-k',-}\left|0\right\rangle\otimes\left(\frac{\sqrt{n+1}}{D_+}\left|n_q+1\right\rangle+\frac{\sqrt{n}}{D_-}\left|n_{-q}-1\right\rangle\right)\\
	\varphi_-=c^+_{k',+}c^+_{-k,-}\left|0\right\rangle\otimes\left(\frac{\sqrt{n+1}}{D_+}\left|n_{-q}+1\right\rangle+\frac{\sqrt{n}}{D_-}\left|n_{q}-1\right\rangle\right)
	\end{array},$$
with $D_{\pm}=\epsilon_k-\epsilon_{k'}\pm\hbar\omega_q$. The searched $x_{k,k'}=N\left\langle k\left|k'_2\right.\right\rangle$ is then inferred to read 
$$x_{k,k'}=\frac{\left(2 n(T)+1\right)g^2_q}{\left(\left(\epsilon_{k}-\epsilon_{k'}\right)^2-\left(\hbar\omega_q\right)^2\right)}\quad , $$
with $n(T)=\left(e^{\frac{\hbar\omega_q}{k_BT}}-1\right)^{-1}$ being the thermal average of $n_{\pm q}$. Moreover it can be checked that $x_{k,k'}=x_{k',k}$. Thus, for $q$ not close to the Brillouin zone center (the most likely occurence), there is $x_{k,k'}>0$, whereas $x_{k,k'}<0$ can be found only for $q\approx 0$. Likewise, though the hereabove expression is redolent of one derived by Froehlich\cite{fro}, their respective significances are unrelated, since Froehlich interpreted the self-energy of \textit{one} electron and  \textit{one} phonon \textit{bound} together in terms of \textit{virtual} transitions between various electron-phonon states, whereas $x_{k,k'}$ refers to the dot product of \textit{two}-electron-states.\par
	Projecting the hermitian BCS Hamiltonian\cite{bcs,ja1,ja2,ja3} $H$ onto the basis $\left\{\left|k_2\right\rangle,\left|k'_2\right\rangle\right\}$ yields
$$\begin{array}{l} 
H_{k_2,k_2}=2\left(\epsilon_k+\frac{x_{k,k'}U}{N^2}+\frac{x^2_{k,k'}}{N^2}\epsilon_{k'}\right)\\
H_{k_2,k'_2}=\frac{U}{N}\left(1+\frac{x^2_{k,k'}}{N^2}\right)+2\frac{x_{k,k'}}{N}\left(\epsilon_k+\epsilon_{k'}\right)\\
H_{k'_2,k'_2}=2\left(\epsilon_{k'}+\frac{x_{k,k'}U}{N^2}+\frac{x^2_{k,k'}}{N^2}\epsilon_{k}\right)
\end{array}\quad ,$$	
whence it can be concluded within the thermodynamic limit ($N\rightarrow\infty$) that the diagonal matrix elements $H_{k,k}$ remains unaltered by the electron-phonon coupling, whereas $U$ is slightly renormalised to $U+2x_{k,k'}\left(\epsilon_k+\epsilon_{k'}\right)$. Anyhow, since, as noted above, $x_{k,k'}>0$ is the most likely case, it is hard to figure out how the phonon-mediated isotope effect could lessen $U$, as concluded by Froehlich\cite{fro}.\par
	Because, in some materials, the observed isotope effect does not comply with $T_c\sqrt{M}=constant$, it has been ascribed tentatively\cite{sab} to the hyperfine\cite{abr} interaction, coupling the nuclear and electron spin, provided the electron wave-function includes some $s$-like character. We shall derive the corresponding $x_{k,k'}$, by proceeding similarly as above for the electron-phonon one and keeping the same notations.\par
	The Hamiltonian reads for nuclear spins $=1/2$ in momentum space
	$$H_h=\frac{A}{\sqrt{N}}\sum_{k,k'}c^+_{k,+}c_{-k',-}I_q^-+c^+_{-k,-}c_{k',+}I_q^+\quad ,$$
with $A$ being the hyperfine constant, $\pm$ referring to the two spin directions and $q=k+k'$. Likewise, the $I^\pm=\frac{\sigma_x\pm i\sigma_y}{2}$'s, with $\sigma_x,\sigma_y$ being Pauli's matrices\cite{abr} characterising the nuclear spin, operate on nuclear spin states $\left|\pm\right\rangle$. Note that the term $\propto \sigma_z$ has been dropped because it turned out to contribute nothing to $x_{k,k'}$. The unperturbed eigenstates read
	$$\left|\widetilde{k}\right\rangle=\left|k\right\rangle\otimes\frac{\left|+\right\rangle_q+\left|-\right\rangle_q}{\sqrt{2}},\quad 
\left|\widetilde{k'}\right\rangle=\left|k'\right\rangle\otimes\frac{\left|+\right\rangle_q+\left|-\right\rangle_q}{\sqrt{2}}\quad .$$
Their respective energies are $E(k)=2\epsilon(k)$, $E(k')=2\epsilon(k')$. Then $x_{k,k'},\left\langle k\left|k'_2\right.\right\rangle$ read in this case
	$$\begin{array}{l}
	x_{k,k'}=-\frac{A^2}{4\left(\epsilon_{k'}-\epsilon_{k}\right)^2}\\
	\left\langle k\left|k'_2\right.\right\rangle=\frac{x_{k,k'}}{N}\left\langle\widetilde{k}
	\left|h_h\right|\varphi\right\rangle\left\langle\varphi\left|h_h\right|\widetilde{k'}\right\rangle\\
	\varphi=c^+_{k,+}c^+_{k',+}\left|0\right\rangle\otimes\left|-\right\rangle_q+c^+_{-k,-}c^+_{-k',-}\left|0\right\rangle\otimes\left|+\right\rangle_q
	\end{array}.$$
Except for having the opposite sign, $x_{k,k'}$ has the same properties as in case of the electron-phonon coupling, which causes $U$ to be renormalised to a slightly lesser value.
				\section{Experimental Outlook}
Three experiments, enabling one to assess the validity of this analysis, will be discussed below. The first one addresses the determination of $\frac{\partial\mu}{\partial c_s}$, which plays a key role for the existence of persistent currents\cite{sz4} and the stability of the superconducting phase (see Eq.(\ref{cri})). As shown elsewhere\cite{sz5}, the partial pressure $p(T\leq T_c)$, exerted by the conduction electrons, and their associated compressibility coefficient\cite{lan3} $\chi(T)$ read
				\begin{equation} 
						\label{com}
				\begin{array}{l}
				p=c_nE_F(c_n)-F_n(c_n)+c_s\mu(c_s)-\mathcal{E}_s(c_s)\Rightarrow\\
				\chi=-\frac{dV}{Vdp}=\left(c_n^2\frac{\partial E_F}{\partial c_n}+c_s^2\frac{\partial\mu}{\partial c_s}\right)^{-1}
				\end{array}\quad ,
				\end{equation}
with $c_n=c_n(T),c_s=c_s(T)$ and $V$ being the sample volume. For $T\rightarrow T_c$, there is $c_s\rightarrow 0$, so that it might be impossible to measure the contribution of bound electrons $\propto c_s^2\frac{\partial\mu}{\partial c_s}(0)$ to $\chi$ in Eq.(\ref{com}). Such a hurdle might be dodged by making the kind of differential measurement to be described now. A square-wave current $I(t+t_p)=I(t),\forall t$, such that $I\left(t\in\left[-\frac{t_p}{2},0\right]\right)=0,I\left(t\in\left[0,\frac{t_p}{2}\right]\right)=I_c$ ($I_c$ stands for the critical current), is flown through the sample, so that the sample switches periodically from superconducting to normal. Then using a lock-in detection procedure for the $\chi$ measurement might enable one to discriminate $c_s^2\frac{\partial\mu}{\partial c_s}$ against $c_n^2\frac{\partial E_F}{\partial c_n}$, despite $c_s\left(T\rightarrow T_c\right)<<c_n\approx c_0$ and thence to check the validity of Eq.(\ref{cri}).\par 
	The validity of Eq.(\ref{gidu}) can be assessed by shining $UV$ light of variable frequency $\omega$ onto the sample and measuring the electron work function\cite{ash} $w(T\leq T_c)$ by observing \textit{two} distinct photoemission thresholds $w_1=\hbar\omega_1=E_F(T),w_2=\hbar\omega_2=2\mu(T)$, associated respectively with single electron and electron pair excitation. Observing $\omega_2=2\omega_1$ would validate Eq.(\ref{gidu}). Besides, if $c_s(T)$ is known from skin-depth measurements\cite{sz1}, $\mu(c_s)$ could be charted. Note also that, if such an experiment were to be carried out in a material, exhibiting a superconducting gap $E_g$, a large decrease of $E_F$ from $E_F(T_c)$ down to $E_F(0)=\mu(c_0)=\epsilon_b-E_g$ should be expected ($\epsilon_b$ designates the bottom of the conduction band).\par
	For $T>10K$, the electron specific heat is overwhelmed\cite{ash} by the lattice contribution $C_\phi$, so that there are no accurate experimental data\cite{lor} for $C_s(T)$. Such a difficulty might be overcome by using again the differential technique, described above. A constant heat power $W$ is fed into a thermally insulated sample, while its time-dependent temperature $T(t)$ is monitored. Thus $T(t)$ can be obtained owing to
	$$\begin{array}{l}
	W=\left(C_\phi(T)+C_s(T)\right)\dot T\left(t\in\left[-\frac{t_p}{2},0\right]\right)\\
	W=\left(C_\phi(T)+C_n(T)\right)\dot T\left(t\in\left[0,\frac{t_p}{2}\right]\right)
	\end{array}\quad,$$
with $\dot T=\frac{dT}{dt}$. Feeding again the square-wave current, mentioned above, into the sample, while using the same lock-in detection technique, could enable one to extract $C_s(T)-C_n(T)$ from the measured signal $\dot T(t)$, despite $C_\phi>>C_s,C_n$. Note\cite{sz5} that $C_\phi,C_n$, unlike $C_s$, do not depend on the current $I$ and $C_n$ can always be measured at low $T$ and then extrapolated\cite{ash} up to $T_c$ thanks to $C_n(T)=\frac{\left(\pi k_B\right)^2}{3}\rho(E_F)T$.
				\section{Conclusion}
A criterion, warranting the stability of the superconducting phase, has been worked out and found to agree with a prerequisite $\frac{\partial\mu}{\partial c_s}<0$, established previously for persistent currents\cite{sz4}, thermal equilibrium\cite{sz5} and the Josephson effect\cite{sz6}. The temperature dependence at $T\rightarrow T_c$ has been given for the specific heat, concentration and maximum current density, conveyed by superconducting electrons. At last, an original derivation of the isotope effect has been given.\par
	Due to the inequality $U\frac{\partial\mu}{\partial c_s}<0$, shown elsewhere\cite{sz5}, the necessary condition $\frac{\partial\mu}{\partial c_s}<0$  entails $U>0$, i.e. a \textit{repulsive} inter-electron force, such as the Coulomb one,  is needed for superconductivity to occur, if the Hubbard model is taken to describe the correlated electron motion. Note that, in the mainstream interpretation\cite{dag,arm} of the properties of high-$T_c$ materials, such a repulsive force is also believed to be instrumental above $T_c$ but \textit{not} below $T_c$ due to the BCS assumption $U<0$, although the nature of the inter-electron coupling remains \textit{unaltered} at $T_c$. Thence, the BCS model is found \textit{not} to be consistent with persistent currents\cite{sz4}, thermal equilibrium\cite{sz5} and a stable superconducting phase, as shown hereabove, due to $U<0\Rightarrow\frac{\partial\mu}{\partial c_s}>0$.

\end{document}